\documentclass[a4paper,10pt,preprint,twocolumn,3p]{elsarticle} 

\usepackage{lineno}
\usepackage{mathtools}
\usepackage{bm,amssymb}
\usepackage{subcaption}
\usepackage{booktabs}
\usepackage{color} 
\usepackage{natbib}
\usepackage{physics}
\usepackage{siunitx}
\usepackage[dvipsnames]{xcolor}
\usepackage[pdftex,colorlinks,citecolor=blue,bookmarks]{hyperref}

\usepackage{comment}

\journal{Physics Letters B}

\biboptions{numbers,sort&compress}
\begin{document}
\onecolumn
\begin{frontmatter}
\title{Neutrinoless $\beta\beta$ decay nuclear matrix elements complete up to N$^2$LO in heavy nuclei}

\author[1,2]{Daniel Castillo \corref{cor1}}
\ead{danicastillog16@gmail.com}

\author[3]{Lotta Jokiniemi \corref{cor1}}
\ead{ljokiniemi@triumf.ca}

\author[1,2]{Pablo Soriano}

\author[1,2]{Javier Men\'{e}ndez}
\ead{menendez@fqa.ub.edu}

\address[1]{%
 Departament de F\'isica Qu\`antica i Astrof\'isica, Universitat de Barcelona, 08028 Barcelona, Spain.
}
\address[2]{
Institut de Ci\`encies del Cosmos, Universitat de Barcelona, 08028 Barcelona, Spain.
}%

\address[3]{TRIUMF, 4004 Wesbrook Mall, Vancouver, BC V6T 2A3, Canada}

\cortext[cor1]{Corresponding author.}

\date{\today}

\begin{abstract}
We evaluate all nuclear matrix elements (NMEs) up to next-to-next-to leading order (N$^2$LO) in chiral effective field theory ($\chi$EFT) for the neutrinoless double-beta ($0\nu\beta\beta$) decay of the nuclei most relevant for experiments, including $^{76}$Ge, $^{100}$Mo, and $^{136}$Xe. We use the proton-neutron quasiparticle random-phase approximation (pnQRPA) and the nuclear shell model to calculate the N$^2$LO NMEs from very low-momentum (ultrasoft) neutrinos and from loop diagrams usually neglected in $0\nu\beta\beta$ studies. Our results indicate that 
the overall N$^2$LO contribution is centered around $-(5$-$10)\%$ for the shell model and $+(10$-$15)\%$ for the pnQRPA, with sizable uncertainties due to the scale dependence of the ultrasoft NMEs and the short-range nature of the loop NMEs. The sign discrepancy between many-body methods is common to all studied nuclei and points to the different behaviour of the intermediate states of the $0\nu\beta\beta$ decay. Within uncertainties, our results for the ultrasoft NME are of similar size as contributions usually referred to as ``beyond the closure approximation''.
\end{abstract}
\end{frontmatter}

\section{\label{sec:intro}Introduction}
Neutrinoless $\beta\beta$ ($0\nu\beta\beta$) decay---a hypothetical process in which two nucleons in an atomic nucleus $\beta$-decay simultaneously without emitting neutrinos---is amongst the most promising probes of physics beyond the standard model of particle physics~\cite{Agostini2023}. Observing $0\nu\beta\beta$ decay would demonstrate lepton-number violation and hence shed light on the matter-antimatter asymmetry of the Universe. It would also show that neutrinos are their own antiparticles, or \emph{Majorana particles}, and give indications on their masses. For all these reasons, the decay is under massive experimental investigations~ \cite{Anton2019,Agostini2020,Majorana:2022udl,Azzolini2022,CUPID-Mo:2023,Adams2024,Abe2024,Agrawal:2024zfv}. While the standard-model-allowed $\beta\beta$ mode with the emission of two neutrinos ($2\nu\beta\beta$) has been observed in a dozen nuclei \cite{Barabash2020}, current experiments only give upper limits to the $0\nu\beta\beta$-decay rates. Next-generation experiments aim at
improving current $0\nu\beta\beta$ sensitivities by two orders of magnitude in the next decade
\cite{Abgrall2021,Adhikari2021,Alfonso2022,NEXT:2020amj}. 

Exploring fundamental physics from $0\nu\beta\beta$ experiments requires the knowledge of the nuclear matrix elements (NMEs) governing the decay. However, they currently form a major obstacle as even the best NME calculations~\cite{Yao:2019rck,Wirth:2021pij,Novario:2020dmr,Belley2023heavy,Belley2024,Weiss:2021rig,Coraggio:2020hwx,Coraggio:2022vgy,Jokiniemi2023c,Horoi:2022ley,Horoi:2023uah,Lv:2023dcy,Simkovic:2018hiq,Mustonen:2013zu,Fang:2018tui,Rodriguez:2010mn,LopezVaquero:2013yji,Ding:2023dnl,Yao:2021wst,Wang2024} carry large  uncertainties \cite{Agostini2023,Gomez-Cadenas:2023vca}. Moreover, a recent $0\nu\beta\beta$ derivation using chiral effective field theory ($\chi$EFT)~\cite{2Cirigliano2018} has pointed out additional uncertainties, showing that the so-called short-range NME, originally proposed at next-to-next-to-leading order (N$^2$LO) in the power counting, had to be promoted to leading order (LO) for the theory to be renormalizable \cite{Cirigliano2018,Cirigliano2019}. Indeed, the LO short-range NME is in the ballpark of a 50\% correction for various many-body methods: the nuclear shell model (NSM) \cite{Jokiniemi2021}, the proton-neutron quasiparticle random-phase approximation  method (pnQRPA) \cite{Jokiniemi2021}, the {\it ab initio} valence-space in-medium similarity renormalization group (VS-IMSRG) approach \cite{Belley2024,Belley2023heavy}, and the generalized contact formalism that combines the NSM with quantum Monte Carlo (QMC) methods \cite{Weiss:2021rig}.

The next $0\nu\beta\beta$ contributions enter at N$^2$LO in the chiral power counting. Some of them are usually taken into account in NME calculations by introducing nucleon form factors and the nucleon isovector magnetic moment operator \cite{Engel2017}. However, the $\chi$EFT analysis reveals that two additional kind of diagrams enter at N$^2$LO~\cite{2Cirigliano2018}. First, a N$^2$LO NME arises from the exchange of virtual ultrasoft neutrinos, with momenta much lower than the Fermi momentum of the nucleus. This is the lowest-order NME where states of the intermediate nucleus of the $\beta\beta$ decay enter explicitly---usually coined as {\it beyond closure approximation}---and has only been studied in the context of $0\nu\beta\beta$ decay with sterile neutrinos for $^{76}$Ge and $^{136}$Xe with the NSM~\cite{Dekens:2023iyc,Dekens2024}. Second, additional N$^2$LO NMEs emerge from loop diagrams. They have only been calculated in light systems with QMC~\cite{Pastore2018}, and in $^{76}$Ge using the VS-IMSRG~\cite{Belley2024}. While previous works suggest that both types of N$^2$LO NMEs have a 5\%-10\% effect, so far there are no consistent, or systematic, studies on their combined impact on the NMEs of different nuclei.

In this Letter, we explore the $0\nu\beta\beta$-decay NMEs up to N$^2$LO in a wide range of $\beta\beta$ emitters, including all nuclei relevant for the most advanced current and future experiments: $^{76}$Ge, $^{82}$Se, $^{100}$Mo,  $^{130}$Te and $^{136}$Xe~\cite{Agostini2023}. We also compare the N$^2$LO contribution from ultrasoft neutrinos with effects usually referred as beyond the closure approximation. Our many-body calculations rely on the pnQRPA and NSM, commonly used to obtain LO NMEs.

\section{\label{sec:doublebeta}Neutrinoless Double-Beta Decay}

Following the $\chi$EFT approach \cite{2Cirigliano2018} up to N$^2$LO, the $0\nu\beta\beta$-decay half-life is
\begin{align}
[t_{1/2}^{0\nu}]^{-1}&=G_{0\nu}\,g_{\rm A}^4\,\frac{m^2_{\beta\beta}}{m_e^2} \nonumber \\
&\times \left|M_{\rm L}^{0\nu}+M_{\rm S}^{0\nu}+M_{\rm usoft}^{0\nu}+M_{\rm loop}^{0\nu}\right|^2\;,
  \label{eq:half-life}
\end{align}
where $G_{0\nu}$ is a phase-space factor for the electrons~\cite{Kotila2012}, 
$g_{\rm A}\simeq 1.27$ is the axial-vector coupling to the nucleon, and $m_{\beta\beta}=\sum_i U_{ei}^2m_i$ (normalized to the electron mass $m_e$), which characterizes the lepton-number violation, depends on the neutrino masses $m_i$ and mixing matrix $U$. Here we focus on the NMEs: the LO long- and short-range, 
$M_{\rm L}^{0\nu}$ and $M_{\rm S}^{0\nu}$ \cite{Jokiniemi2021}, and the N$^2$LO terms: $M_{\rm usoft}^{0\nu}$  coming from ultrasoft neutrinos  and $M_{\rm loop}^{0\nu}$ resulting from loop diagrams that cannot be absorbed into the nucleon form factors entering $M^{0\nu}_{\rm L}$. 

The standard long-range NME can be written in the familiar way~\cite{Engel2017}
\begin{equation}
M_{\rm L}^{0\nu}=M_{\rm GT}^{0\nu}-\Big(\frac{g_{\rm V}}{g_{\rm A}}\Big)^2M_{\rm F}^{0\nu}+M_{\rm T}^{0\nu}\;,
\label{eq:m_0v}
\end{equation}
with vector coupling $g_{\rm V}=1.0$ and Gamow-Teller (GT, ${\bm \sigma}_m{\bm \sigma}_n$),  Fermi (F, $\bm{1}_{mn}$) and tensor (T, $3[(\boldsymbol{\sigma}_m\cdot\hat{\mathbf{r}})(\boldsymbol{\sigma}_n\cdot\hat{\mathbf{r}})]-\boldsymbol{\sigma}_m\cdot\boldsymbol{\sigma}_n=S_{mn}^{\rm T}$) parts according to the operator form in spin (${\boldsymbol{\sigma}}$) space, with ${\mathbf{r}}=\mathbf{r}_m-\mathbf{r}_n$.
In $\chi$EFT, the dominant GT term reads~\cite{2Cirigliano2018}
\begin{align}
\label{eq:Mgt}
  & M^{0\nu}_\text{GT} = \frac{2R}{\pi g_{\rm A}^2} \times \\
&\langle 0^+_f||
\sum_{m,n}
\tau^-_m\tau^-_n\,{\bm \sigma}_m{\bm \sigma}_n
\int j_0(qr)h_\text{GT}(q^2) {\rm d}q|| 0^+_i\rangle, \nonumber
\end{align}
evaluated between the initial $0^+_i$ and final $0^+_f$ states summing over all $A$ nucleons' spin and isospin ($\tau$) operators, with momentum transfer $q$, Bessel function $j_0$, and $R=1.2 A^{1/3}\,$fm. This expression agrees with the one obtained from a phenomenological approach in the 
closure approximation (with no closure energy)~\cite{Engel2017}. 
$M^{0\nu}_\text{GT}$ also depends on a neutrino potential, with $h_\text{GT}(0)=g_{\rm A}^2$ and additional $q$-dependent terms, including partial N$^2$LO corrections regularized with dipole form factors as in previous pnQRPA~\cite{Hyvarinen2015,Simkovic2008} and NSM~\cite{Menendez18} studies.
The Fermi and tensor parts
follow similar expressions to~Eq.~(\ref{eq:Mgt}).
Finally, we correct our many-body states with two-nucleon short-range correlations (SRCs) following the so-called 
CD-Bonn and Argonne parametrizations~\cite{Simkovic2009}, even though other works suggest that SRCs may have a larger impact~\cite{Weiss:2021rig,Benhar:2014cka}.

The LO short-range NME
connects directly the initial and final nuclei~\cite{Cirigliano2018}
\begin{align}\label{eq:contact}
  M^{0\nu}_{\rm S} = \frac{2R}{\pi g_{\rm A}^2}
\langle 0^+_f||
\sum_{m,n}
\tau^-_m\tau^-_n\,\!\!
\int \!j_0(qr)h_{\rm S}(q^2)\,q^2 {\rm d}q|| 0^+_i\rangle,
\end{align}
regularized with a Gaussian regulator of scale $\Lambda$:
\begin{equation}
h_{\rm S}(q^2)=2g_{\nu}^{\rm NN}\,e^{-q^2/(2\Lambda^2)}\;.
\label{eq:h-contact}
\end{equation}
We approximate the unknown coupling $g_{\nu}^{\rm NN}$ with the charge-independence-breaking terms of different nuclear Hamiltonians like in \cite{Jokiniemi2021} with their consistent cutoffs that vary between $\Lambda=349 \cdots 550$~MeV.

The N$^2$LO ultrasoft NME depends on the $1^+_n$ states of the intermediate nucleus, with energies $E_n$. The NME expression is given by \cite{2Cirigliano2018, Dekens2024}
\begin{align}
\label{eq:Musoft}
M_{\rm usoft}^{0\nu}&=\!\sum_n\langle 0_f^+||\!\sum_k\! \tau_k^-\boldsymbol{\sigma}_k||1_n^+\rangle \langle 1_n^+||\!\sum_k\! \tau_k^-\boldsymbol{\sigma}_k||0_i^+\rangle \nonumber \\
&\times -\frac{2R}{\pi} \left(\frac{Q_{\beta\beta}}{2}+m_e+E_n-E_i\right) \nonumber \\
&\times\left[\ln\left(\frac{\mu_{\rm us}}{2\left(\frac{Q_{\beta\beta}}{2}+m_e+E_n-E_i\right)}\right)+1\right]\!,
\end{align}
where $Q_{\beta\beta}$ is the $\beta\beta$-decay $Q$-value, and $E_i$ the energy of the initial state. Here we assume that the emitted electrons share equally the energy release: $E_{e_1}=E_{e_2}=Q_{\beta\beta}/2+m_e$, which only introduces a very small error~\cite{Dekens2024}.

Finally, the N$^2$LO loop NME can be decomposed into four terms~\cite{2Cirigliano2018}
\begin{align}
    M^{0\nu}_{\rm loop}&= 
    M^{0\nu}_{\rm loop,VV}+M^{0\nu}_{\rm loop,AA}
    +M^{0\nu}_{\rm loop,CT} \nonumber \\ &
    +    M^{0\nu}_{\rm loop, usoft}\;,
\end{align}
with each part ($X$=VV, AA, CT, usoft) given by
\begin{equation}
     M^{0\nu}_{{\rm loop},X}=\frac{4R}{\pi g_{\rm A}^2}\langle 0^+_f||\sum_{m,n}
\tau^-_m\tau^-_n\mathcal{V}^{(m,n)}_{  X}||0^+_i\rangle\;.
\end{equation}
The vector (VV) part reads 
\begin{align}
    &\mathcal{V}^{(m,n)}_{\rm VV}=\int {\rm d}q\; q^2e^{-q^2/(2\Lambda^2)}\frac{g^2_A\hat{q}}{3(4\pi F_{\pi})^2}
    \nonumber 
    \\
    &\times\left(\frac{2(1-\hat{q})^2}{\hat{q}^2(1+\hat{q})}\text{ln}(1+\hat{q})-\frac{2}{\hat{q}}+\frac{7-3\hat{q}L_{\pi}}{(1+\hat{q})^2}+\frac{L_{\pi}}{1+\hat{q}}\right)
    \nonumber 
    \\
    &\times\left(j_0(qr)\boldsymbol{\sigma}_m\cdot\boldsymbol{\sigma}_n-j_2(qr)S_{mn}^{\rm T}\right)\;,
    \label{eq:n2lo_vector}    
\end{align}    
the axial-vector (AA) part is
\begin{align}
    &\mathcal{V}^{(m,n)}_{\rm AA}=\int {\rm d}q\; q^2e^{-q^2/(2\Lambda^2)}\bigg[\frac{g^2_A\hat{q}}{3(4\pi F_{\pi})^2} 
    \nonumber
    \\
    &\times\left(\frac{g^2_A}{1+\hat{q}}(4-L_{\pi})-\frac{1}{(1+\hat{q})^2}\right)
    \nonumber
    \\
    &\times\left(j_0(qr)\boldsymbol{\sigma}_m\cdot\boldsymbol{\sigma}_n-j_2(qr)S_{mn}^{\rm T}\right)
    \nonumber
    \\
    &+\frac{1}{(4\pi F_{\pi})^2}\Bigg(\frac{3}{4}(1-g^2_A)^2L_{\pi}-g^4_Af_4(\hat{q})-g^2_Af_2(\hat{q})
    \nonumber
    \\
    &-f_0(\hat{q})-24g^2_AF^2_{\pi}C_T(L_{\pi}+1)\Bigg)j_0(qr)\bigg]
    \;,
\label{eq:n2lo_axial}
\end{align}    
the counterterm (CT)---that should absorb the renormalization scale ($\mu$) dependence of the VV, AA, and usoft terms---is given by
\begin{equation}
\begin{split}
    &\mathcal{V}^{(m,n)}_{\rm CT}=
    \int {\rm d}q\; q^2e^{-q^2/(2\Lambda^2)}\frac{g^2_A\hat{q}}{3(4\pi F_{\pi})^2}\\
    &\times\Big(-\frac{5}{6}g^{\pi\pi}_{\nu}\frac{\hat{q}}{(1+\hat{q})^2}+g^{\pi N}_{\nu}\frac{1}{1+\hat{q}}\Big)\\
    &\times(j_0(qr)\boldsymbol{\sigma}_m\cdot\boldsymbol{\sigma}_n-j_2(qr)S_{mn}^{\rm T})\;,
    \end{split}
    \label{eq:n2lo_CT}    
\end{equation}
and the ultrasoft term---which, if one uses the leading-order $\chi$EFT Hamiltonian, cancels the dependence on the ultrasoft scale ($\mu_{\rm us}$) of $M_{\rm usoft}^{0\nu}$---is defined as
\begin{equation}
    \begin{split}
       &\mathcal{V}^{(m,n)}_{\rm usoft}=
    \int {\rm d}q\; q^2e^{-q^2/(2\Lambda^2)}\bigg[-\frac{2g^4_A}{3(4\pi F_\pi)^2}\frac{{q}^2}{{q}^2+m_\pi^2}\\
    &\times(j_0(qr)\boldsymbol{\sigma}_m\cdot\boldsymbol{\sigma}_n-j_2(qr)S_{mn}^{\rm T})\\
    &-\frac{2g^4_A}{(4\pi F_{\pi})^2}\frac{{q^2}}{{q^2}+m^2_{\pi}}+\frac{g^2_A}{(4\pi)^2}48C_T\bigg] \ln\left(\frac{m^2_\pi}{\mu_\text{us}^2}\right)\;.\\
    \end{split}
\end{equation}
We vary the ultrasoft scale range as $\mu_\text{us} = m_\pi/2\dots2m_\pi$, given that $M^{0\nu}_{\rm loop, usoft}$ is proportional to $\text{ln}(m^2_\pi/\mu_\text{us}^2)$. We use the short-hand notations $\hat{q}=q^2/m_{\pi}^2$, $L_{\pi}=\ln (\mu^2/m_{\pi}^2)$, the $f_n(\hat{q})$ functions from \cite{2Cirigliano2018}, $C_T=\pi/m_{\rm N}(1/(a_t^{-1}-\mu)-1/(a_s^{-1}-\mu))$ with $a_s^{-1}=-8.3$~MeV and $a_t^{-1}=36$~MeV,  $g_{\nu}^{\pi\pi}=-10.84-36/5\ln(\mu/m_\rho)$ from the average of the lattice QCD evaluations \cite{Tuo2019,Detmold2020}, $F_{\pi}=92.28$~MeV and $g_{\nu}^{\pi N}$=0. Since the constants $C_T$ and $g_{\nu}^{\pi\pi}$ are evaluated around the $\rho$ mass, we vary the scale $\mu$ in the range $\mu=500\cdots 1500$~MeV as in \cite{Belley2024}. We regularize the loop NMEs with the same Gaussian regulators as the LO short-range NME.

In this work, we give results combining the two N$^2$LO terms that depend on $\mu_{\rm us}$, which we refer to as {\it total ultrasoft} NME, $M_{\rm tot, usoft}^{0\nu}$. On the other hand, we label the sum of the remaining loop NMEs as {\it loop soft}, $M_{\rm loop, soft}^{0\nu}$, so that
\begin{align}
 M_{\rm usoft}^{0\nu}+&M_{\rm loop}^{0\nu}
= 
\left(M_{\rm usoft}^{0\nu}+M_{\rm loop, usoft}^{0\nu}\right) \nonumber \\
&+\left(M^{0\nu}_{\rm loop,VV}+M^{0\nu}_{\rm loop,AA}+M^{0\nu}_{\rm loop,CT}\right) \nonumber \\
  &= M_{\rm tot, usoft}^{0\nu}+ M_{\rm loop, soft}^{0\nu}\,.
\end{align}

\begin{table*}[h]
    \centering
    \caption{Central values and ranges of all NMEs up to N$^2$LO: LO long range (second and sixth columns), LO short range (third and seventh), N$^2$LO total ultrasoft (fourth and eighth) and N$^2$LO soft loops (fifth and ninth). The NME ranges cover two interactions for the NSM or a range of $g_{pp}$ values for the pnQRPA, as well as various couplings $g_{\nu}^{\rm NN}$, Gaussian regulators $\Lambda=349\cdots 550$ MeV, CD-Bonn and Argonne SRCs, the renormalization scale range $\mu=500\cdots 1500$ MeV, and the ultrasoft scale $\mu_{\rm us}=m_\pi/2\cdots2m_\pi$ (see text for details).}
    \begin{tabular}{c*{8}{S[table-format=2.1(3)]}}
    \toprule
    Nucleus &\multicolumn{4}{c}{pnQRPA} &\multicolumn{4}{c}{NSM}\\
    \cmidrule(lr){2-5}\cmidrule(lr){6-9}
    &\multicolumn{2}{c}{LO} &\multicolumn{2}{c}{N$^2$LO} &\multicolumn{2}{c}{LO} &\multicolumn{2}{c}{N$^2$LO}\\
    \cmidrule(lr){2-3}\cmidrule(lr){4-5}\cmidrule(lr){6-7}\cmidrule(lr){8-9}
    &L &S &tot~usoft & loop~soft &L &S &tot~usoft &loop~soft\\
    \midrule
    $^{48}$Ca & & & & &0.92(14) &0.43(20) & 0.01(6) 
    & 0.05(7)\\
    $^{76}$Ge &5.23(58) &2.65(116) &0.39(31) &0.28(41) &3.57(25) &0.97(48) & -0.29(9)
    &0.05(16)\\
    $^{82}$Se &4.56(36) &2.26(99) &0.30(19) &0.21(34) &3.38(20) &0.91(43) &-0.27(8)
    &0.05(15)\\
    $^{96}$Zr &4.28(24) &2.22(98) &0.49(23) &0.25(33)\\
    $^{100}$Mo &3.44(73) &2.96(130) &1.15(52) &0.38(43)\\ 
    $^{116}$Cd &4.85(38) &1.95(85) &-0.12(3) &0.15(28)\\ 
    $^{124}$Sn &5.20(32) &2.99(130) &0.59(35) &0.37(46) &2.79(63) &1.06(52) &-0.23(13)
    &0.06(16)\\
    $^{130}$Te &3.70(34) &2.12(94) &0.46(38) &0.31(34) &2.68(79) &1.07(50) &-0.23(15)
    &0.06(16)\\
    $^{136}$Xe &2.99(28) &1.36(60) &0.06(14) &0.16(32) &2.26(53) &0.86(41) &-0.19(11)
    &0.05(13)\\
         \bottomrule
    \end{tabular}
    \label{tab:ranges}
\end{table*}

\section{\label{sec:many-body}Many-Body NME Calculations}

We perform NSM calculations with the codes ANTOINE and NATHAN~\cite{Caurier2005}. We use three different valence spaces, always symmetric for neutrons and protons: the $pf$-shell ($0f_{7/2}$, $1p_{3/2}$, $0f_{5/2}$ and $1p_{1/2}$ orbitals) on top of a $^{40}$Ca core for $A=48$, where we use the KB3G~\cite{Poves:2000nw} and GXPF1A~\cite{Honma2002} interactions; 
the $1p_{3/2}$, $0f_{5/2}$, $1p_{1/2}$ and $0g_{9/2}$ orbitals with a $^{56}$Ni core for $A=76$ and $A=82$, where we use the GCN2850~\cite{Menendez2009} and JUN45~\cite{Honma2009} interactions; and for $A=124$, $A=130$ and $A=136$ we use the $0g_{7/2}$, $1d_{5/2}$, $1d_{5/2}$, $2s_{1/2}$ and $0h_{11/2}$ orbitals  with the GCN5082~\cite{Caurier:2010az} and QX~\cite{Qi2012} interactions. Overall, with the NSM we study six $0\nu\beta\beta$ decays: $^{48}$Ca, $^{76}$Ge, $^{82}$Se, $^{124}$Sn, $^{130}$Te, and $^{136}$Xe. 
We consider all nuclear configurations in the full valence space except in $^{124}$Te (final nucleus of the $^{124}$Sn $\beta\beta$ decay)
which is limited to seniority $v\leq5$ states (up to five broken zero-angular-momentum pairs) instead of the full $v\leq6$. We have checked that, nonetheless, the NMEs are converged to the percent level. 

We also compute the NMEs with the spherical pnQRPA method \cite{Suhonen2007}. We use the same no-core single-particle bases as in \cite{Jokiniemi2021,Jokiniemi2023}: they consist of 18 orbitals for $A=76,82$ nuclei, 25 orbitals for $A=96,100$, and 26 orbitals for $A=124-136$. In this framework we study the decays of $^{76}$Ge, $^{82}$Se, $^{96}$Zr, $^{100}$Mo, $^{116}$Cd, $^{124}$Sn, $^{130}$Te, and $^{136}$Xe.
We take the single-particle energies from a Coulomb-corrected Woods-Saxon potential optimized for nuclei close to $\beta$ stability~\cite{Bohr1969} and slightly modified in the vicinity of the Fermi surfaces to better reproduce the low-lying spectra of neighboring odd-mass nuclei.
The pnQRPA is based on quasiparticle spectra for protons and neutrons, which are solved from Bardeen-Cooper-Schrieffer equations with the  interaction derived from the Bonn-A potential \cite{Holinde1981} and proton and neutron pairing parameters fine-tuned to reproduce the empirical pairing gaps. The residual Hamiltonian contains two  more parameters: the particle-hole ($g_{\rm ph}$)  and particle-particle ($g_{\rm pp}$) pairing strengths.
We fix $g_{\rm ph}$ to reproduce the centroids of the GT giant resonance, 
and $g_{\rm pp}$ to the $2\nu\beta\beta$-decay half-life according to the partial isospin-symmetry restoration scheme~\cite{Simkovic2013}, using both the bare $g_{\rm A}$ and an effective $g_{\rm A}^{\rm eff}=1.0$. 
For $^{96}$Zr and $^{124}$Sn, adjusting $g_{\rm pp}$ to $2\nu\beta\beta$-decay is not possible, so we use a fixed value below the pnQRPA breaking point. For details, see \cite{Jokiniemi2023}.

\section{\label{sec:results}Results and Discussion}
\subsection{\label{ssec:musoft-impact}Ultrasoft NMEs and closure approximation}

Table~\ref{tab:ranges} presents the results of all NMEs up to N$^2$LO calculated with the NSM and pnQRPA. The central values and ranges cover the interactions, couplings, regulators and SRCs discussed in Secs.~\ref{sec:doublebeta} and ~\ref{sec:many-body}.
Focusing on the central values, the results in Table~\ref{tab:ranges} indicate that for the NSM the total ultrasoft NME is about $5\%$-$10\%$ and has opposite sign compared to the total LO NME---summing the long- and short-range parts. The only exception is $^{48}$Ca, where the ultrasoft NME is suppressed. In contrast, for the pnQRPA the total ultrasoft NME has the same sign and $5\%$-$10\%$ of the value of the LO NME. Nonetheless, in $^{100}$Mo it reaches $40\%$ and it is very small in $^{116}$Cd and $^{136}$Xe. 
The uncertainty in $M_{\rm tot, usoft}^{0\nu}$ is significant, reaching $50\%$ for several nuclei in the NSM and pnQRPA calculations. This error is a measure of the inconsistency for not using a $\chi$EFT Hamiltonian, as it stems mainly from the $\mu_{\rm us}$ dependence of the NSM and pnQRPA NMEs. For the NSM the error in $M^{0\nu}_{\rm usoft}$ dominates, whereas for the pnQRPA the uncertainties for $M^{0\nu}_{\rm usoft}$ and $M^{0\nu}_{\rm loop, usoft}$ are comparable.

\begin{figure}[t!]
    \centering
    \includegraphics[width=\linewidth]{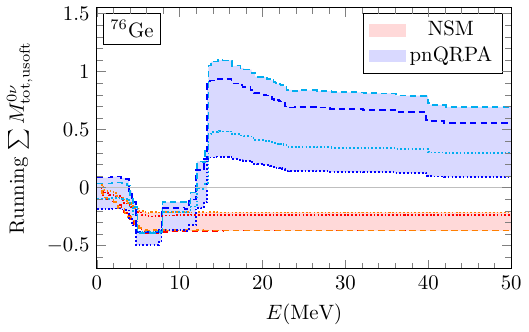}
    \caption{Running sum of the total ultrasoft NME of $^{76}$Ge, $M_{\rm tot, usoft}^{0\nu}$, as a function of the excitation energy of the intermediate $1^+$ states, $E$, for the NSM and pnQRPA. The NSM results include the JUN45 (orange) and GCN2850 (red) interactions, while the pnQRPA results comprise $g_{\rm pp}$ values adjusted to $2\nu\beta\beta$-decay with $g_{\rm A}^{\rm eff}=1.0$ (blue) and $g_{\rm A}=1.27$ (cyan). The ranges cover the variation of the scale $\mu_{\rm us}$ from $m_{\pi}/2$ (dotted) to $2m_{\pi}$ (dashed).}
    \label{fig:musoft_running_sums}
\end{figure}

In order to investigate the opposite sign of the ultrasoft NME in NSM and pnQRPA calculations, Fig.~\ref{fig:musoft_running_sums} compares the running sums of the $M_{\rm tot, usoft}^{0\nu}$ in $^{76}$Ge as a function of the excitation energy of the intermediate states, $E$, for both many-body methods. The results  cover the NSM interactions and pnQRPA $g_{\rm pp}$ range presented in Sec.~\ref{sec:many-body}.
Figure~\ref{fig:musoft_running_sums} shows that up to $E\approx 8$ MeV the two running sums are rather similar. However, they differ markedly around $8\text{ MeV}\lesssim E\lesssim13\text{ MeV}$, where large cancellations and a resulting sign change appear in the pnQRPA but are not present in the NSM results. In fact, this behaviour is common to all the studied nuclei with the pnQRPA, while in the NSM we only observe $M_{\rm tot, usoft}^{0\nu}$ cancellations, although milder, in $^{48}$Ca. Similar differences between running sums in the NSM and pnQRPA have been noticed in the context of $2\nu\beta\beta$ decay \cite{Gando2019}---which involves a similar sum over $1^+$ intermediate states. While the NSM may not capture these high-energy contributions because of missing spin-orbit partners in the valence spaces used, the pnQRPA may overestimate them. Measurements of charge-exchange reactions up to the GT resonance energy in both $\beta^-$ and $\beta^+$ directions, currently limited to lower energies~\cite{Fujita:2011za,Frekers:2018edj}, could solve this discrepancy.

\begin{figure}[t!]
    \centering
    \includegraphics[width=\linewidth]{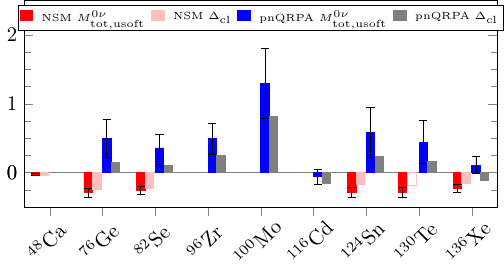}
    \caption{Difference between non-closure and LO long-range NMEs ($\Delta_\text{cl}$, pink and gray bars) compared with the total ultrasoft NME ($M_{\rm tot, usoft}^{0\nu}$, red and dark blue bars) obtained by the NSM and pnQRPA for all decays studied in this work. The NSM results use closure energies from \cite{Sen'kov2016,Sarkar2024,Sarkar2024b} (solid bars) except for $^{130}$Te where we average the closure energies of $^{124}$Sn and $^{136}$Xe (open bar). Results for Argonne SRCs.}
    \label{fig:Closure-nonclosure}
\end{figure}

In $\chi$EFT \cite{2Cirigliano2018}, $M^{0\nu}_{\rm usoft}$ can be seen as a correction beyond closure~\cite{Rodin:2006bbw,Sen'kov2013,Sen'kov2014,Sen'kov2014b,Sen'kov2016,Sarkar2024,Sarkar2024b}, since it is the lowest-order NME that explicitly depends on the structure of the intermediate states. 
We compare the difference of $M^{0\nu}_{\rm L}$ calculated with and without closure approximation, $\Delta_{\rm cl}\equiv M^{0\nu}_{\rm non-cl} - M^{0\nu}_{\rm L}$, with $M^{0\nu}_{\rm tot, usoft}$, which takes into account the uncertainty due to $\mu_\text{us}$ in $M^{0\nu}_{\rm usoft}$ and also in the related term $M^{0\nu}_{\rm loop, usoft}$. This uncertainty, reflected in the error bars, would vanish in a fully consistent calculation. We compute $\Delta_{\rm cl}$ in the pnQRPA and use the results of \cite{Sen'kov2016,Sarkar2024,Sarkar2024b} for the NSM---for $^{130}$Te we average between the closure energies for $^{136}$Xe \cite{Sarkar2024} and $^{124}$Sn \cite{Sarkar2024b}. Figure \ref{fig:Closure-nonclosure} shows the comparison. 
Considering the uncertainties,  $M^{0\nu}_{\rm tot, usoft}$ is consistent with beyond-closure effects in all nuclei for the NSM.
In contrast, for the pnQRPA, $M^{0\nu}_{\rm tot, usoft}$ amounts to roughly twice $\Delta_{\rm cl}$ in all nuclei, except for $^{136}$Xe where the two quantities are small. Nonetheless, the sign of both quantities is consistently the same.

\subsection{\label{ssec:N2LO-impact} N$^2$LO loop NMEs}

Table~\ref{tab:ranges} presents the calculated central values and ranges of the N$^2$LO loop NMEs. While the central values are typically small, less than $5\%$ of the total LO NME for both many-body methods, the uncertainties are significant. When included, $M^{0\nu}_{\rm loop, soft}$ can reach up to $10\%$ of the total LO NME for the pnQRPA---{$15\%$} for $^{100}$Mo---, and up to $5\%$ for the NSM, which are roughly the relative effects of the ultrasoft NMEs.

\begin{figure}[t]
    \centering
    \includegraphics[width=\linewidth]{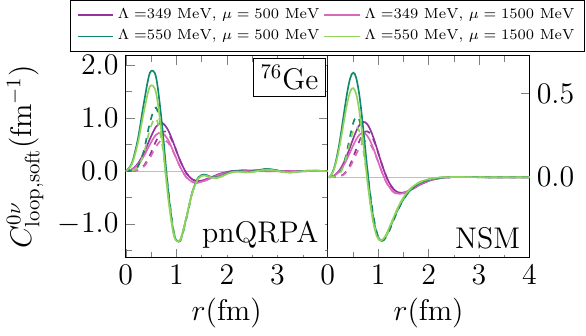}
    \caption{Radial distributions of the N$^2$LO loop NME for $^{76}$Ge, $C^{0\nu}_{\rm loop, soft}$, obtained with the pnQRPA (left) and NSM (right). The results cover Gaussian regulators $\Lambda=349$~MeV (purple) and $\Lambda=550$~MeV (green), renormalization scales $\mu=500$~MeV (dark colors) and $\mu=1500$~MeV (light), ultrasoft scale $\mu_\text{us}=m_\pi$, and Argonne (dashed lines) and CD-Bonn (solid) SRCs.}
    \label{fig:n2lo_radial_densities}
\end{figure}

To illuminate the origin of the large uncertainties for $M^{0\nu}_{\rm loop, soft}$, Fig.~\ref{fig:n2lo_radial_densities} shows, for the NSM and pnQRPA, their radial distribution for $^{76}$Ge, which satisfies $M^{0\nu}_{\rm loop, soft}=\int_{0}^{\infty}C^{0\nu}_{\rm loop, soft}(r){\rm d}r$. 
Different curves show the results for different regulators $\Lambda$, scales $\mu$, and SRCs.
Qualitatively, the radial distributions look very similar in the pnQRPA (left) and NSM (right), but the overall scales are quite different. This is consistent with the comparison of the LO NMEs~\cite{Jokiniemi2021}. 
Figure \ref{fig:n2lo_radial_densities} also indicates that the main sensitivity of the radial distribution of $M^{0\nu}_{\rm loop, soft}$ is to the regulator $\Lambda$ (difference between the blue and red curves: choosing a dipole regulator like in \cite{Pastore2018} gives results in between these two results). This sensitivity stems from the short-range---high-momentum---nature of the loop NME, also highlighted in Fig.~\ref{fig:n2lo_radial_densities}, which concentrates at short-distances almost like the LO short-range NME~\cite{Jokiniemi2021}. This makes the N$^2$LO loop NME very sensitive to SRCs as well: choosing either CD-Bonn or Argonne parameterizations can even change the sign of $M^{0\nu}_{\rm loop, soft}$. In contrast, the dependence on the scale $\mu$ is rather mild. 

\begin{table*}[ht!]
    \centering
    \caption{Results for the LO long-range (second and eighth columns), LO short-range (third and ninth), N$^2$LO ultrasoft (fourth and tenth), and N$^2$LO loop VV (fifth and eleventh), AA (sixth and twelfth) and CT (seventh and thirteenth) NMEs for all nuclei studied in this work. In the pnQRPA, we fix $g_{\rm pp}$ with bare $g_{\rm A}$, and in the NSM we use the KB3G, GCN2850 and GCN5082 interactions. The results correspond to Argonne SRCs, short-range coupling $g_{\nu}^{\rm NN}=-1.03$ fm$^2$, Gaussian regulator  $\Lambda=349$ MeV, and scales $\mu=m_{\rho}$ 
    and $\mu_{\rm us}=m_{\pi}$.
    }
    \begin{tabular}{c*{12}{S[table-format=2.1]}}
    \toprule
    Nucleus &\multicolumn{6}{c}{pnQRPA} &\multicolumn{6}{c}{NSM}\\
    \cmidrule(lr){2-7}\cmidrule(lr){8-13}
    &\multicolumn{2}{c}{LO} &\multicolumn{4}{c}{N$^2$LO} &\multicolumn{2}{c}{LO} &\multicolumn{4}{c}{N$^2$LO}\\
    \cmidrule(lr){2-3}\cmidrule(lr){4-7}\cmidrule(lr){8-9}\cmidrule(lr){10-13}
    &L &S &usoft &VV &AA &CT &L &S &usoft &VV &AA &CT\\
    \midrule
         $^{48}$Ca& & & & & & & 0.96 & 0.37 & -0.03 
         & -0.07 &0.03 &0.11 \\
         $^{76}$Ge&4.68 &1.93 &0.50 &-0.40 &0.16 &0.68 & 3.32 & 0.85 & -0.29 
         & -0.11 &0.05& 0.18 \\
         $^{82}$Se&4.20 &1.65 &0.36 &-0.33 &0.13 &0.55 & 3.18 & 0.80 & -0.28 
         & -0.10 &0.05 & 0.17 \\
         $^{96}$Zr&4.04 &1.60 &0.49 &-0.33 &0.14 &0.56 & & & &\\
         $^{100}$Mo&2.71 &2.14 &1.30 &-0.47 &0.19 &0.82 & & & &\\
        $^{116}$Cd&4.47 &1.47 &-0.06 &-0.25 &0.10 &0.43 & & & &\\
        $^{124}$Sn&4.88 &2.18 &0.58 &-0.49 &0.19 &0.85&3.17 & 0.90& -0.28 
        &-0.11 &0.05 &0.24\\ 
         $^{130}$Te&3.36 &1.52 &0.44 &-0.39 &0.15 &0.67 & 3.24 & 0.94 & -0.29
         & -0.12 & 0.06 & 0.19 \\
         $^{136}$Xe&2.71 &0.99 &0.11 &-0.23 &0.09 &0.40 & 2.60 & 0.75 & -0.23
         & -0.09 &0.04 & 0.15 \\
         \bottomrule
    \end{tabular}
    \label{tab:all_corrections}
\end{table*}

Table \ref{tab:all_corrections} gives a decomposition of the VV, AA and CT components of the N$^2$LO loops NME for a particular combination of nuclear interaction, couplings, regulators and SRCs---common for the NSM and pnQRPA results. The LO and N$^2$LO ultrasoft NMEs obtained with the same combination of parameters are also listed for comparison. The results for the components of $M^{0\nu}_{\rm loop, soft}$ indicate that, both for the NSM and pnQRPA, the different parts cancel, with the AA and CT terms sharing the same sign as the LO NMEs, and VV carrying systematically the opposite sign. Individually, each of the three N$^2$LO loops contributions of soft neutrinos can be of similar size, or even larger than the N$^2$LO ultrasoft NME; however, due to the cancellation, the total $M^{0\nu}_{\rm loop, soft}$ tends to be smaller than $M^{0\nu}_{\rm tot, usoft}$, see Table~\ref{tab:ranges}.  

\subsection{\label{ssec:powercounting}Combined N$^2$LO NMEs}

\begin{figure}
    \centering
    \includegraphics[width=\linewidth]{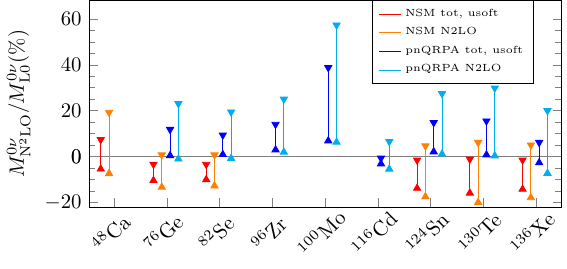}
    \caption{Relative contributions of the total ultrasoft ($M^{0\nu}_{\rm tot, usoft}$, red and dark blue bars) and complete N$^2$LO (orange and light blue) NMEs compared to the LO NME, obtained with the NSM and pnQRPA for all nuclei studied in this work. The bars show the uncertainties in Table \ref{tab:ranges}.}
    \label{fig:NME_total}
\end{figure}
With the results in Table \ref{tab:all_corrections}, we can compare the relative size of the $0\nu\beta\beta$ NMEs. In particular, the largest N$^2$LO contributions are typically the ultrasoft NME for the NSM, and the CT part of the loops NME for the pnQRPA. They amount to $\lesssim10\%$ and $\lesssim20\%$ of the standard long-range LO NME, respectively, which is somewhat larger than expected from the $\chi$EFT power counting~\cite{2Cirigliano2018,Dekens2024}. The ultrasoft N$^2$LO NME is especially large for $^{100}$Mo---about 50\% of the LO long-range NME. However, even in this extreme case the combined N$^2$LO NME is about a third of the total LO NME.

The size of the largest N$^2$LO NMEs is nonetheless similar to the impact of the nucleon dipole form factors used for  $M^{0\nu}_{\rm long}$, typically a $10\%$-$15\%$ effect~\cite{Menendez2009,Rodin:2006bbw}, from diagrams that enter mainly at the same order. The magnetic moment operator, a N$^2$LO term as well, also contributes at the $10\%$ level for several nuclei~\cite{Hyvarinen2015,Menendez18}. In addition, in Sec.~\ref{ssec:musoft-impact} we discussed that, for the NSM, the relatively large ultrasoft NMEs are of similar size as beyond closure effects.

Figure \ref{fig:NME_total} summarizes the ratios of the N$^2$LO and total LO NMEs, with the uncertainty ranges from Table \ref{tab:ranges}. The relative impact of $M^{0\nu}_{\rm tot, usoft}$ is typically a $5\%$-$10\%$ enhancement (pnQRPA, blue bars) or reduction (NSM, red bars). Likewise, the central values of the combined N$^2$LO contributions also differ in sign, even though the NSM errorbars (orange) allow for a positive contribution as predicted by the pnQRPA (light blue). The central values of the total N$^2$LO NMEs are about $-(5$-$10)\%$ for the NSM and $+(10$-$15)\%$ for the pnQRPA, with the extreme absolute values being limited to
$\leq 20 \%$ and $\leq 30 \%$, respectively, except in the extreme case of $^{100}$Mo where it can reach up to $60\%$ of the LO NME.

\section{\label{sec:conclusions}Summary and outlook}

We have calculated for the first time complete $0\nu\beta\beta$ NMEs up to N$^2$LO for nine medium-mass and heavy nuclei including the most promising experimental candidates $^{76}$Ge, $^{100}$Mo, $^{130}$Te and $^{136}$Xe. 

The absolute central value of the ultrasoft NMEs ($5\%$-$10\%$ of the total LO NME) are typically larger than the NMEs coming from loop diagrams that cannot be absorbed into nuclear form factors. Both terms introduce sizable uncertainties, due to their ultrasoft-scale dependence or short-range character. When considering theoretical uncertainties, ultrasoft NMEs are of similar size and same sign as beyond-closure effects.

The main discrepancy between the NSM and pnQRPA N$^2$LO NMEs is that for the former $M^{0\nu}_{\rm tot, usoft}$ has opposite sign to the LO NMEs, while for the latter it has the same sign. This is due to the different behaviour of the states in the intermediate $0\nu\beta\beta$ nucleus, which could be probed experimentally in charge-exchange reactions. The difference between the results of the two many-body methods persists for the combined N$^2$LO NMEs, which typically range between $-20\%$ to $+5\%$ in the NSM, and $0\%$ to $+30\%$ for the pnQRPA. Our N$^2$LO NMEs are anomalously large for $^{100}$Mo: the ultrasoft term can be up to 40\% and the combined N$^2$LO contribution up to 60\%. Overall, N$^2$LO NMEs should be included in any $0\nu\beta\beta$ calculation since their impact is moderate but comparable to the one of other N$^2$LO diagrams, like nucleon form factors, routinely included in NME studies.

For the future, we would like to narrow down the uncertainties in $M^{0\nu}_{\rm loop}$ by investigating more carefully the dependence with different regulators and regulator forms. We would also like to use $\chi$EFT Hamiltonians to check that the $\mu_\text{us}$ scale dependence of $M^{0\nu}_{\rm tot, usoft}$ vanishes. Additionally, we would like to include two-body currents, which in the $\chi$EFT power counting enter at N$^3$LO \cite{2Cirigliano2018,Pastore2018}. Nonetheless, in $\beta$ decays the importance of two-body currents is critical to understand the overestimation of matrix elements in many-body calculations (or ``quenching'')~\cite{Gysbers:2019uyb,Coraggio:2023eep}. In fact, based on the impact on $\beta$ decays, some estimations for $0\nu\beta\beta$ decay suggest an effect comparable to the N$^2$LO NMEs studied in this work~\cite{Menendez2011, Engel2014,Jokiniemi2023c}, even though smaller contributions are also possible~\cite{Engel2018}.

\section*{Acknowledgements}
We would like to thank A. Belley, V. Cirigliano, W. Dekens, M. Drissi, and J. Engel for insightful discussions. We thank the Institute for Nuclear Theory at the University of Washington for its kind hospitality and stimulating research environment. This research was supported in part by the INT's U.S. Department of Energy grant No. DE-FG02- 00ER41132. This work was supported by Arthur B. McDonald Canadian Astroparticle Physics Research Institute, the
MCIN/AEI/10.13039/5011
00011033 from the following grants: PID2020-118758GB-I00, PID2023-147112NB-C22, RYC-2017-22781 through the “Ram\'on
y Cajal” program funded by FSE “El FSE invierte en tu futuro”, CNS2022-135716 funded by the
“European Union NextGenerationEU/PRTR”, and CEX2019-000918-M to the “Unit of Excellence Mar\'ia de Maeztu 2020-2023” award to the Institute of Cosmos Sciences; and by the Generalitat de Catalunya, grant 2021SGR01095.
TRIUMF receives federal funding via a contribution agreement with the National Research Council of Canada. 

\bibliography{contact-refs}

\end{document}